\documentstyle[prl,multicol,aps,psfig]{revtex} % for camera-ready compuscript

\begin{document}
\title{ 
Comment on ``Critical temperature of trapped hard-sphere Bose gases''
}
\author{Werner Krauth}
\address{
CNRS-Laboratoire de Physique Statistique de l'ENS\\
24, rue Lhomond; F-75231 Paris Cedex 05; France\\
e-mail: krauth@physique.ens.fr\\
}
\date{December 1998}
\maketitle
\begin{abstract}
In this comment, I discuss a recent path-integral Monte Carlo 
calculation by Pearson, Pang, and Chen ( {\sl Phys. Rev. A}
{\bf 58}, 4796 (1998)). For bosons with a small hard-core interaction
in a harmonic trap, the authors find a critical temperature
which does not change with respect to the non-interacting gas. 
The calculation suffers from a serious discretization error
of the many-particle density matrix.
\pacs{PACS numbers: 02.70Lq 05.30Jp }
\end{abstract}
\begin{multicols}{2}
\narrowtext

In a recent article, Pearson, Pang, and Chen \cite{Pearson} studied
hard-core bosons in a harmonic trap using Path-Integral Quantum
Monte Carlo (PIMC) methods. The authors conclude that the critical
temperature of this system for small hard-core diameters $a$ does
not notably change with respect to the non-interacting gas. These
findings are difficult to reconcile both with previous Monte Carlo
work \cite{Krauth} and with mean-field calculations \cite{Giorgini},
which agree very nicely with each other. However, the calculation
of Pearson, Pang, and Chen is flawed by a discretization error
which affects the many-particle density matrix for small $a$.

In \cite{Pearson}, the density matrix $\rho(R,R',\beta)$ at inverse
temperature $\beta$ is reduced to an integral over high-temperature
matrices $\rho(R,R',\tau)$ with $\tau = \beta/M$, and  $M$ the
number of `slices'  
\begin{equation}
\rho(R,R',\beta) =\int d R_2 \cdots
\int  d R_M \rho(R,R_2,\tau)  \cdots  \rho(R_M,R' ,\tau)
\label{convolution}
\end{equation}
({\em cf} \cite{Pearson} for notation, $R=(r_1,\ldots,r_N)$ is the
$3 N$-dimensional vector of particle coordinates, with $N$ the
number of particles; I use natural units with mass $m=1$, oscillator
length $a_0=1$, $\hbar=1$).

To evaluate   $\rho(R_i,R_{i+1},\tau)$, the authors
use the so-called ``primitive approximation'' ({\em cf} \cite{Ceperley})
$\rho(R_i,R_{i+1} ,\tau) \sim \rho_P(R_i,R_{i+1} ,\tau)$ with 
\begin{equation}
\rho_P(R_i,R_{i+1},\tau) = c   
\exp(-\frac{ (R_i-R_{i+1})^2}{2 \tau } -\tau 
V(R_i)  )  
\label{primitive} 
\end{equation}
For the present discussion, the harmonic potential is unessential
and I will only consider the hard-core term.
 
The interaction $V(r)$ sets a small length scale, $a$,
and the primitive approximation is justified only if, between $R_i$
and $R_{i+1}$, the paths fluctuate by much less than this scale. 
Under this condition we can conclude that, if paths overlap
neither at slice $i$ nor at slice $i+1$, they will not collide in
between. The spatial fluctuations are determined by the kinetic energy 
term in Eq. (\ref{primitive}), and one arrives at the  condition
\begin{equation}
   \tau \sim  (r_i - r'_i)^2  \ll  a^2
\label{estimate} 
\end{equation}
In their work, Pearson, Pang and Chen consider, {\em e. g.} a value
$a=0.02$. It follows from Eq. (\ref{estimate}) that $\tau$ must
satisfy $\tau \ll  0.0004$. For a typical value  of $\beta=0.2$,
this implies $M \gg  500$.  In ref. \cite{Pearson}, $M \sim 10$ is
used.  The effect of the interaction is therefore largely {\em
under}estimated and important deviations from the case of
noninteracting bosons are not picked up.

The condition $\tau \ll  a^2$ renders the use of the primitive
approximation unpractical for large systems of weakly interacting
bosons. However, it is very instructive to check the estimate Eq.
(\ref{estimate}) for a {\em  single pair} of particles in the
setting of ref. \cite{Pearson}.  To do this, I have generated a
large number of $3-$dimensional Gaussian paths directly from the
L\'{e}vy construction ({\em cf} \cite{Ceperley}) for various values of
$M$.  These paths sample Eq. (\ref{convolution})  for $V=0$.
$\rho(R,R',\beta)$ is obtained by means of a correction factor,
which differs from unity by the probability $P$ of pairs of paths
to collide at some slice $i$.
\begin{figure}
\centerline{ \psfig{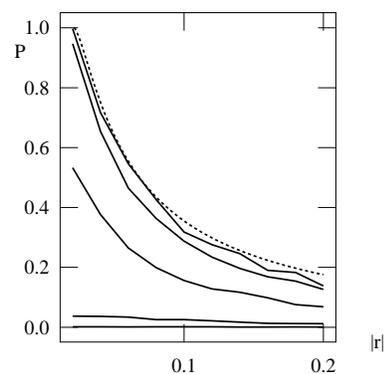} }
\caption{Collision probability $P$ for a pair
of particles with hard-core diameter $a=0.02$ as a function of their
separation $|r|$ (in reduced units, $a_0=1$, $\beta=0.2$), 
for $M=10,100,\ldots,100,000 $ (full lines, from below).
Also shown is $P$ in the limit $M \rightarrow 
\infty$ obtained from the eigenfunctions of the hard-sphere potential 
(dotted line).}
\end{figure}
As an example, I consider the diagonal density matrix for two
particles separated by a distance $|r|$ ($R=(0,r)$, $R'=(0,r)$).
In fig. 1, the numerically determined probability $P$ is plotted
for $\beta=0.2$ as a function of $|r|$ for values of $\tau =0.02$
($M=10$) to $\tau =2\times  10^{-6}$ ($M=100,000$).  As predicted
in Eq. (\ref{estimate}), we need large values of $M$ to obtain a
converged result. In the limit $M\rightarrow \infty$, $P$ can also
be computed directly from the eigenfunctions of the hard-sphere
potential ({\em cf} \cite{Krauth}), as presented in fig. 1.  This
shows that the interaction of two particles need not be calculated
with Monte Carlo methods. This computation is at the heart of the
recent successful PIMC calculations for weakly interacting bosons,
both in the trap \cite{Krauth} \cite{Holzmann} and in free space
\cite{Grueter}.

\noindent
\end{multicols}
\end{document}